\documentclass[aps,twocolumn,pra,floatfix,longbibliography,superscriptaddress]{revtex4-2}
\pdfoutput=1
\usepackage{graphicx,braket,float}
\usepackage{amsmath}
\usepackage{amsfonts}
\usepackage{amsbsy}
\usepackage{mathtools}
\usepackage{xcolor}
\usepackage{soul}
\usepackage[normalem]{ulem}
\usepackage[unicode]{hyperref}
\usepackage[export]{adjustbox}
\usepackage{enumitem}
\hypersetup{
    unicode=true, % non-Latin characters in Acrobat?s bookmarks
    plainpages=false,
    colorlinks=true,% false: boxed links; true: colored links
    linkcolor=blue,% color of internal links
    citecolor=blue,% color of links to bibliography
    filecolor=blue,% color of file links
    urlcolor=blue% color of external links
}
\begin{document}

\title{Extracting work from coherence in a two-mode Bose-Einstein condensate}
\author{L. A. Williamson}
\thanks{Corresponding author: lewis.williamson@uq.edu.au}
\affiliation{ARC Centre of Excellence for Engineered Quantum Systems, School of Mathematics and Physics, University of Queensland, St Lucia, Queensland 4072, Australia}
\author{F. Cerisola}
\affiliation{Department of Physics and Astronomy, University of Exeter, Stocker Road, Exeter EX4 4QL, United Kingdom}
\author{J. Anders}
\affiliation{Department of Physics and Astronomy, University of Exeter, Stocker Road, Exeter EX4 4QL, United Kingdom}
\affiliation{Institut f\"{u}r Physik und Astronomie, University of Potsdam, 14476 Potsdam, Germany}
\author{Matthew J. Davis}
\thanks{mdavis@uq.edu.au}
\affiliation{ARC Centre of Excellence for Engineered Quantum Systems, School of Mathematics and Physics, University of Queensland, St Lucia, Queensland 4072, Australia}
\date{\today}

\begin{abstract}
We show how work can be extracted from number-state coherence in a two-mode Bose-Einstein condensate. With careful tuning of parameters, a sequence of thermodynamically reversible steps transforms a Glauber coherent state into a thermal state with the same energy probability distribution. The work extracted during this process arises entirely from the removal of quantum coherence. More generally, we characterise quantum (from coherence) and classical (remaining) contributions to work output, and find that in this system the quantum contribution can be dominant over a broad range of parameters. The proportion of quantum work output can be further enhanced by squeezing the initial state. Due to the many-body nature of the system, the work from coherence can equivalently be understood as work from entanglement.
\end{abstract}

\maketitle

\section{Introduction}
Converting disordered energy (heat) into ordered energy (work) is one of the most fundamental processes in thermodynamics. In a classical system, disorder arises from practical limitations on what can be known about a large system. Observables in a quantum system may exhibit not only classical uncertainty, but also quantum uncertainty arising from coherence~\cite{aberg2006,streltsov2017}. Quantum coherence is defined with respect to a particular basis, and occurs when a system exists in superpositions of eigenstates of that basis. Of particular relevance to thermodynamics is the quantum uncertainty in a system's energy, arising from coherence in the energy eigenbasis~\cite{vinjanampathy2016,goold2016,binder2018}.

It is now well established that coherence in the energy eigenbasis can enhance work extraction if appropriately utilised. For a sufficiently rapid engine cycle, coherence can enhance power output beyond that of any classical engine operating with the same resources~\cite{uzdin2015,uzdin2016,klatzow2019}. In single-shot realisations, coherence may diminish the maximum available work~\cite{horodecki2013}, however, a carefully chosen sequence of thermodynamic processes circumvents this degradation~\cite{scully2003,kammerlander2016,korzekwa2016,francica2020}. Despite the benefits of coherence in work extraction, protocols to realise these benefits in experimentally tractable systems are lacking. Such expositions are timely, considering the relevance to quantum information~\cite{brandao2013,chitambar2019,ciampini2017} and the growing field of quantum thermal machines~\cite{millen2016,cangemi2023}.

Bose-Einstein condensates (BECs) offer a pristine system in which to explore many-body quantum physics, due to a high degree of tuneability and isolation from the environment~\cite{bloch2012,langen2015}. In the field of quantum thermal machines, bosonic particle statistics can improve engine performance~\cite{myers2022,jaramillo2016,fogarty2020,simmons2023} and act as a fuel in an engine-like cycle~\cite{koch2023}. Interactions in a BEC can enhance engine performance~\cite{bengtsson2018,boubakour2023} or allow access to energetic degrees of freedom not available classically~\cite{chen2019,watson2023}. However, the role of coherence in work extraction from a BEC has not been explored. Notably, two-mode BECs can exhibit long-lived coherence in the number state basis~\cite{jo2007}. Incorporating squeezing~\cite{orzel2001,sorensen2001,esteve2008}, two-mode BECs can exhibit metrologically useful entanglement, allowing for quantum-enhanced measurement sensitivity~\cite{muessel2014,pezze2018,szigeti2020}.

In this manuscript we show how number-state coherence in a two-mode BEC can empower work extraction. To obtain work that is predominantly quantum in origin the initial state must be close to thermal when projected onto the energy eigenbasis. This is challenging in a quantum many-body system due to the exponentially large number of populations and coherences. We show that this challenge can be overcome by using a Glauber coherent state with mean boson number less than a critical value that depends on temperature. We introduce a formalism to quantify the work directly available from coherence and the work that would be available classically, and show that squeezing the initial state can substantially increase the proportion of quantum work extracted. The number-state coherence in our protocol arises from entangling correlations between the two modes of the BEC. Therefore the work extracted from coherence can equivalently be interpreted as work from entanglement, thus demonstrating the many-body nature of the protocol. To our knowledge, our work presents the first proposal showing how work can be extracted from coherence in a two-mode BEC, paving the way for experimental demonstration and highlighting the utility of two-mode BECs in thermodynamic applications.

\section{Background}
\subsection{Extracting work from coherence}\label{extractwork}
To see how work can be extracted from quantum coherence generally, consider a Hamiltonian $\hat{H}$ with thermal state $\rho_\mathrm{therm}=e^{-\beta \hat{H}}/\operatorname{Tr}e^{-\beta \hat{H}}$ at inverse temperature $\beta$. Now consider a second quantum state $\rho$ that satisfies
\begin{equation}\label{En}
\braket{E_n|\rho|E_n}=\braket{E_n|\rho_\mathrm{therm}|E_n},
\end{equation}
with $\ket{E_n}$ the energy eigenstates of $\hat{H}$. The information entropy associated with energy measurements of state $\rho$ (termed ``energy entropy'' in~\cite{binder2018}) is
\begin{equation}\label{Se}
S^E=-\sum_n \braket{E_n|\rho|E_n}\ln \braket{E_n|\rho|E_n},
\end{equation}
which for an isolated system is non-decreasing under time evolution~\cite{polkovnikov2011,santos2011}. When Eq.~\eqref{En} is satisfied, the energy entropy of state $\rho$ is identical to that of $\rho_\mathrm{therm}$. If $\rho$ is diagonal in the energy eigenbasis then $S^E$ is equal to the von Neumann entropy $S^V=-\operatorname{Tr}\left[\rho\ln\rho\right]$ and $\rho$ is equal to $\rho_\mathrm{therm}$, which is a completely passive state~\cite{pusz1978}. If, however, $\rho$ contains coherences (off-diagonal) terms in the energy eigenbasis, $S^E$ will differ from $S^V$~\cite{polkovnikov2011,kosloff2017,balian1989}. The difference between the two,
\begin{equation}\label{SeSv}
S^E-S^V=\sum_{m,n\ne m} \braket{E_m|\rho|E_n}\braket{E_n|\ln\rho|E_m},
\end{equation}
can be extracted in the form of non-classical work~\cite{kammerlander2016,francica2020,esposito2011}
\begin{equation}\label{Wrelen}
W_\mathrm{quant}=\beta^{-1}(S^E-S^V)=\beta^{-1}D(\rho ||\rho_\mathrm{ed}).
\end{equation}
Here $D(\rho ||\sigma)=\operatorname{tr}[\rho\ln\rho]-\operatorname{tr}[\rho\ln\sigma]\ge 0$ is the quantum relative entropy, and $\rho_\mathrm{ed}=\sum_n \braket{E_n|\rho|E_n}\ket{E_n}\bra{E_n}$ is the {\it projection} of the density matrix $\rho$ onto the energy eigenstates (the ``energy diagonal'' density matrix). The quantity $D(\rho ||\rho_\mathrm{ed})$ is a monotonic measure of the quantum coherence of state $\rho$~\cite{baumgratz2014} and hence the work output, Eq.~\eqref{Wrelen}, is quantum in origin.

A protocol to extract work from coherence was presented in~\cite{kammerlander2016}. The efficacy of this protocol depends on how closely the system's mean energy and energy entropy match that of the thermal state $\rho_\mathrm{therm}$. When these match, any work extracted during the protocol is entirely from coherence. A sufficient condition for this matching is that $\rho_\mathrm{ed}=\rho_\mathrm{therm}$, Eq.~\eqref{En}. This can always be satisfied in a two-level system by choosing the temperature of $\rho_\mathrm{therm}$ to satisfy~\cite{quan2007}
\begin{equation}
k_B T=\frac{E_1-E_0}{\ln\frac{\braket{E_0|\rho|E_0}}{\braket{E_1|\rho|E_1}}}.
\end{equation}
Here $E_0$ and $E_1$ are the two energy levels with $E_1>E_0$.

In many-body systems, it is usually much harder to engineer experimentally a non-thermal state $\rho$ satisfying $\rho_\mathrm{ed}=\rho_\mathrm{therm}$. Achieving this requires
\begin{equation}\label{rn}
r_n=\frac{E_n-E_{n-1}}{\ln\frac{\braket{E_{n-1}|\rho|E_{n-1}}}{\braket{E_n|\rho|E_n}}},
\end{equation}
to be insensitive to $n$, in which case $k_BT=r_n$~\cite{quan2007,plastina2014}. Engineering this for non-thermal many-body states $\rho$ is often very difficult, since control of individual energy levels is usually not possible. As a result, the work output from coherence from such systems will be small. As we will show below, the energy-level structure of an interacting two-mode BEC allows $\rho_\mathrm{ed}\approx \rho_\mathrm{therm}$ for particular experimentally realisable initial states, hence allowing for work output with a high contribution from coherence.

\subsection{System setup}
We consider a two-mode BEC with fixed total boson number $N$. The two modes could be hyperfine spin states or spatial modes~\cite{leggett2001}. Number states of the system are denoted by $\ket{n,k}$, with $n$ and $k=N-n$ the number of bosons in the two modes respectively. We denote the annihilation operators for the two modes by $\hat{a}$ and $\hat{b}$, with $\hat{a}\ket{n,k}=\sqrt{n}\ket{n-1,k}$ and $\hat{b}\ket{n,k}=\sqrt{k}\ket{n,k-1}$. The Hamiltonian for the system is ($\hbar\equiv 1$ here and throughout)~\cite{pezze2018}
\begin{equation}\label{H}
\begin{split}
\hat{H}=&\omega_1 \hat{a}^\dagger\hat{a}+\omega_2\hat{b}^\dagger\hat{b}+\frac{g_{11}}{2}\hat{a}^\dagger\hat{a}^\dagger\hat{a}\hat{a}+\frac{g_{22}}{2}\hat{b}^\dagger\hat{b}^\dagger\hat{b}\hat{b}\\
&+g_{12}\hat{a}^\dagger\hat{b}^\dagger\hat{a}\hat{b}+E_0\\
\equiv &\Delta \hat{a}^\dagger \hat{a}+g \hat{a}^\dagger \hat{a}^\dagger \hat{a}\hat{a}.
\end{split}
\end{equation}
Here $\omega_i$ is the single-particle energy and $g_{ii}$ the interaction strength for modes $i=1,2$. We have also allowed for interactions between the two modes of strength $g_{12}$, for example if the two modes are spin states. The vacuum energy is $E_0$. The operator $\hat{N}=\hat{a}^\dagger\hat{a}+\hat{b}^\dagger\hat{b}$ commutes with $\hat{H}$ and hence the total boson number is conserved. In the second line of Eq.~\eqref{H} we replace $\hat{b}^\dagger\hat{b}$ by $\hat{N}-\hat{a}^\dagger\hat{a}$ and define
\begin{equation}
\Delta\equiv \omega_1-\omega_2-(N-1)(g_{22}-g_{12})
\end{equation}
and
\begin{equation}
g\equiv \frac{g_{11}+g_{22}-2g_{12}}{2}.  
\end{equation}
We choose $E_0=-N\omega_2-N(N-1)g_{22}/2$ so that $\hat{H}\ket{0,N}=0$ for convenience. The choice of $E_0$ will not affect the net output of any cyclic process, for example the performance of an engine cycle. The detuning $\Delta$ for two spatial modes can be controlled via tuning the trapping potential~\cite{albiez2005}, and for two spin states via tuning of a Zeeman field~\cite{matthews1998,hall1998}. As will be discussed below, we require thermal states that have a Gaussian (or approximately Gaussian) distribution over states $\ket{n,N-n}$, which requires $g>0$. Hence the interaction strengths must satisfy $g_{11}+g_{22}>2g_{12}$, i.e.\ the average intramode interaction strength $(g_{11}+g_{22})/2$ must exceed the intermode interaction strength $g_{12}$. For two spatially separated modes we have $g_{12}=0$ and therefore $g>0$ is ensured as long as the intramode interactions $g_{ii}$ are positive, as realised for example in $^{87}$Rb and $^{23}$Na condensates~\cite{anderson1995,davis1995}.

Particle conservation confines the system to the subset of states $\ket{n,N-n}$. Since number states are eigenstates of Eq.~\eqref{H}, the number-state coherence realisable in a two-mode BEC~\cite{jo2007} is equivalent to the energy coherence enabling work extraction, Eq.~\eqref{SeSv}. The thermal state of $\hat{H}$ is,
\begin{equation}\label{rhoT}
\rho_\mathrm{therm}=\frac{1}{Z}e^{-\beta \hat{H}}=\sum_{n=0}^N p_n^\mathrm{therm}\ket{n,N-n}\bra{n,N-n},
\end{equation}
with
\begin{equation}\label{pnT}
p_n^\mathrm{therm}=\frac{1}{Z}e^{-\beta (gn(n-1)+\Delta n)}
\end{equation}
the thermal occupation of state $\ket{n,N-n}$ and
\begin{equation}\label{partition}
Z=\sum_{n=0}^N e^{-\beta (gn(n-1)+\Delta n)}
\end{equation}
the partition function. Due to particle interactions, the boson number distribution of a thermal state is approximately Gaussian (when $\beta g>0$ and $-\beta\Delta\gg \sqrt{\beta g}$), in contrast to the Planck distribution of non-interacting bosons. This feature will be essential for the efficacy of our protocol as it will enable close matching between the initial and thermal boson number distributions. We assume $\beta g\ll 1$; noting that $gN\sim \mu$, where $\mu$ is the BEC chemical potential, this is ensured when $\beta \mu/N\ll 1$, a condition satisfied in most realisations of dilute gaseous BECs~\cite{ketterle1999,dalfovo1999}. We can then approximate the sum in Eq.~\eqref{partition} by an integral. We also assume $N\gg 1$ and $\langle\hat{a}^\dagger\hat{a}\rangle\ll N$, i.e.\ most of the particles are in the $\hat{b}$ mode throughout the protocol. We can then set $N\rightarrow\infty$ and approximate $\ket{n,N-n}$ by a harmonic oscillator number state, which simplifies our calculations~\cite{radcliffe1971,holstein1940}. Evaluating Eq.~\eqref{partition} with these two approximations gives
\begin{equation}\label{partitionIntegral}
Z\approx \sqrt{\frac{\pi}{\beta g}}e^{\beta (g-\Delta)^2/(4g)}\left(\frac{1+\operatorname{erf}\left(\frac{\beta(g-\Delta)}{2\sqrt{\beta g}}\right)}{2}\right).
\end{equation}
We denote the $N\rightarrow\infty $ approximation of $\ket{n,N-n}$ by $\ket{n}$ below, however it should be kept in mind that this state is an implicitly two-mode state.

\section{Results}
Our objective is to find an experimentally realisable density matrix containing coherence in the energy eigenbasis with the same (or almost the same) energy and energy entropy as a thermal state Eq.~\eqref{pnT}. As we will now argue, a Glauber coherent state (GCS) satisfies these criteria. A GCS is,
\begin{equation}\label{rhoC}
\ket{\alpha}=\hat{D}(\alpha)\ket{0}=\sum_{n=0}^\infty c_n\ket{n},
\end{equation}
with $\hat{D}(\alpha)=e^{\alpha \hat{a}^\dagger -\alpha^*\hat{a}}$ the displacement operator and $c_n=e^{-|\alpha|^2/2}\frac{\alpha^n}{\sqrt{n!}}$~\cite{glauber1963b}. Such a state can be realised via coherent rotation of a pure state $\ket{0,N}$ or $\ket{N,0}$~\cite{milburn1997,raghavan1999,albiez2005,gati2007,matthews1999,zibold2010} or via interference measurements~\cite{javanainen1996,cirac1996,castin1997} (see also~\cite{barnett1996}), and gives rise to phase coherence between the two modes of the BEC~\cite{andrews1997,hall1998b,kasevich2002,albiez2005,carruthers1965,buzek1992}. Note the upper limit of the sum in Eq.~\eqref{rhoC} is set to $\infty$ by assuming $|\alpha|^2\ll N$ and $N\gg 1$ (see the discussion above Eq.~\eqref{partitionIntegral}). A GCS is a superposition of energy eigenstates of Eq.~\eqref{H} and hence $\ket{\alpha}\bra{\alpha}$ contains off-diagonal terms $c_m^*c_n\ket{n}\bra{m}$ ($m\ne n$); hence a GCS exhibits coherence with respect to the energy eigenbasis. The GCS number-state distribution $p_n^G=|c_n|^2$ is Poissonian with mean and variance equal to $|\alpha|^2$. For large $|\alpha|^2$ we have,
\begin{equation}\label{png}
p_n^G=\frac{|\alpha|^{2n}e^{-|\alpha|^2}}{n!}\approx \frac{1}{\sqrt{2\pi |\alpha|^2}}e^{-(n-|\alpha|^2)^2/(2|\alpha|^2)}.
\end{equation}
Choosing
\begin{equation}\label{tuningparams}
|\alpha|^2=\frac{g-\Delta}{2g},\hspace{1cm}\beta g=\frac{1}{2|\alpha|^2},
\end{equation}
we obtain $p_n^G\approx p_n^\mathrm{therm}$, with $p_n^\mathrm{therm}$ given by Eq.~\eqref{pnT}. Equivalently, it is easy to show that Eq.~\eqref{rn} is independent of $n$ when $|\alpha|^2$ is chosen as in Eq.~\eqref{tuningparams}. Hence with precise tuning of parameters, as given in Eq.~\eqref{tuningparams}, work can be extracted from coherence as discussed in Sec.~\ref{extractwork}.

For more general parameters, the work output will have both a contribution from coherence and a ``classical'' contribution. Below we present an explicit protocol to show how work can be extracted from the coherence of a GCS. We then show how the work output depends on particle number and temperature, identifying a regime where the work output is dominated by the contribution from coherence. Next we show how the work output can be improved by squeezing the initial state and finally we discuss the role of entanglement in the system.

\subsection{Protocol to extract work from a GCS}
\begin{figure*}
\includegraphics[trim=0cm 9cm 0cm 9cm,clip=true,width=0.8\textwidth]{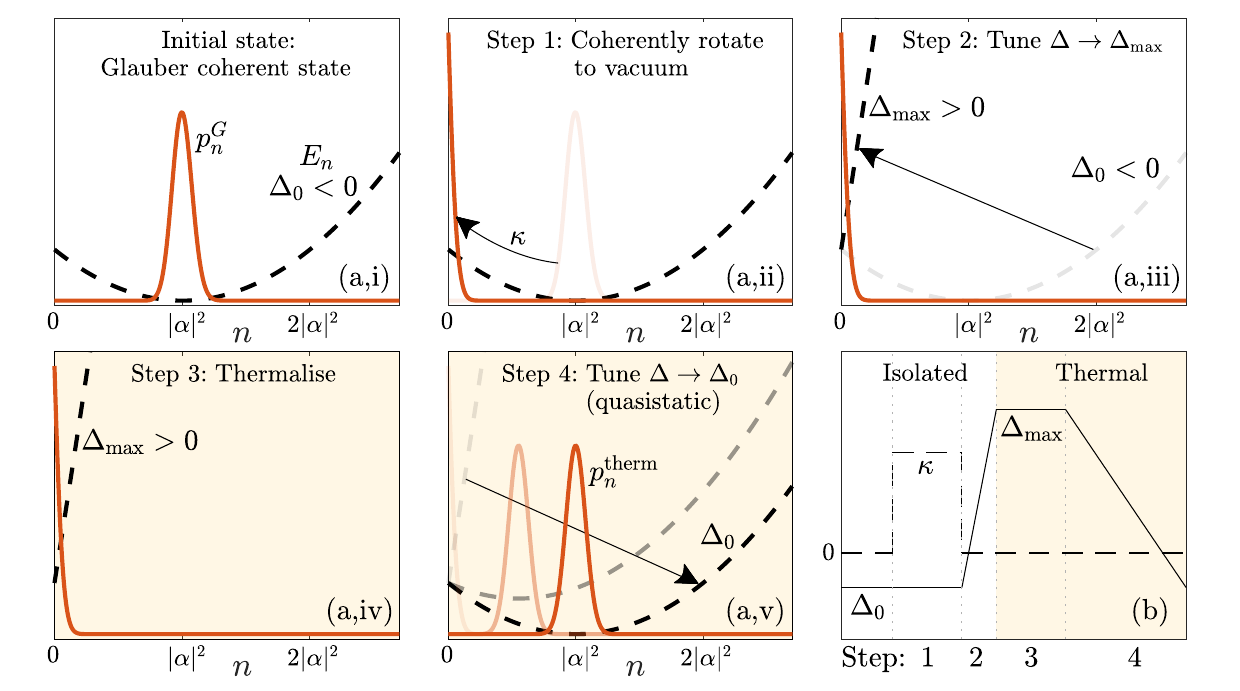}
\caption{\label{schematic} (a) Protocol to extract work from coherence in a two-mode BEC.  Red curves are number state distributions $\braket{n|\rho|n}$ and black dashed curves are the energy spectrum $E_n$, both smoothed over the discrete variable $n$. (a,i) The system is initialised in a Glauber coherent state (GCS) at detuning $\Delta_0$ with mean boson number $|\alpha|^2\gg 1$ and number distribution $p_n^G$, Eq.~\eqref{png}. The energy spectrum has a minimum at $n_\mathrm{min}=\operatorname{max}(0,(g-\Delta_0)/2)$; we choose $\Delta_0<0$ in the schematic above so that $n_\mathrm{min}>0$. (a,ii) Step 1: Work is exchanged as the system is rotated to the vacuum state via a coherent coupling $\kappa$. (a,iii) Step 2: The detuning is changed to $\Delta_\mathrm{max}\gg \beta^{-1}$. The new energy spectrum has a minimum at $n=0$ with an energy spacing much larger than $\beta^{-1}$. Hence the vacuum state is approximately thermal. (a,iv) Step 3: The system is coupled to a reservoir at inverse temperature $\beta$ (indicated by background colouring). (a,v) Step 4: Work is extracted isothermally as the detuning is quasistatically adjusted to its initial value $\Delta_0$. The closeness of the final thermal distribution $p_n^\mathrm{therm}$ to the initial distribution $p_n^G$ determines the quantum and classical contributions to the total work extracted in steps 1-4. The classical contribution can be suppressed with careful tuning of parameters, in which case the work extracted arises entirely from the initial energy coherence. (b) Detuning (solid line) and the coherent coupling (dashed line) between the two modes of the BEC during the steps in (a). }
\end{figure*}

We begin in a GCS, Eq.~\eqref{rhoC}, with Hamiltonian Eq.~\eqref{H} and $\Delta=\Delta_0$. The energy of the initial state is $U_i=\braket{\alpha|\hat{H}|\alpha}=\Delta_0|\alpha|^2+g|\alpha|^4$. To extract work from the GCS we consider the following thermodynamically reversible steps~\cite{kammerlander2016} (see Fig.~\ref{schematic}):
\begin{enumerate}

\item Coherently couple the two modes of the BEC via the Hamiltonian $\kappa(\hat{b}^\dagger\hat{a}+\hat{a}^\dagger\hat{b})$, with $\kappa>0$ the coupling rate. Evolve the system so that $\ket{\alpha}$ is displaced to $\ket{0}$ and all particles are in the $\hat{b}$ mode. During this step, coherence is removed and work $U_i$ is exchanged with the coupling field over a time scale $\kappa^{-1}$. Turn off the coupling $\kappa$ so that the system energy is determined by Eq.~\eqref{H}.

\item Tune the detuning to $\Delta_\mathrm{max}>0$. This does not change the system energy since the system is in state $\ket{0}$.

\item Couple the system to a reservoir at inverse temperature $\beta$ and allow to thermalise. The detuning $\Delta_\mathrm{max}$ and $\beta$ are chosen to satisfy $\beta\Delta_\mathrm{max}\gg 1$. Occupation of modes $\ket{n}$ with $n>0$ are therefore thermally suppressed (see Eq.~\eqref{partition}) and hence $\ket{0}$ is approximately the thermal state.  The energy increase during thermalisation is $\approx \Delta_\mathrm{max} e^{-\beta\Delta_\mathrm{max}}$ and is hence negligible for $\beta\Delta_\mathrm{max}\gg 1$.

\item Quasistatically change the detuning back to $\Delta_0$ such that the system remains in thermal equilibrium with the reservoir throughout. Work is extracted and heat is absorbed in this step. The final state is thermal with energy $U_f=-\partial \ln Z/\partial\beta$ evaluated from Eq.~\eqref{partitionIntegral} with $\Delta=\Delta_0$.

\end{enumerate}
The coherent coupling in the first step could be achieved using Josephson oscillations for two spatial modes~\cite{albiez2005,gati2007} or via a Rabi pulse for two spin states~\cite{matthews1999,zibold2010}. Complete conversion is possible when $\kappa$ is much larger than the other energy scales in the problem, which ensures the dynamics is in the Josephson regime, rather than the self-trapping regime~\cite{milburn1997,raghavan1999,leggett2001,simon2012}. Note the effect of changing $\Delta$ in steps 2-4 could also be achieved by manipulating $g$ using a Feshbach resonance~\cite{papp2008,tojo2010,chin2010} with $\Delta$ fixed: $g$ would be increased to a value $\beta g_\mathrm{max}\gg 1$ in steps 2 and 3 and then quasistatically reduced back to its initial value in step 4. This alternative protocol would not change the results below.

The total mean work output from the protocol in Fig.~\ref{schematic} is~\cite{kammerlander2016}
\begin{equation}
W=U_i-U_f-\beta^{-1}(S^V_i-S^V_f),
\end{equation}
where $S_i^V (S_f^V)$ are the initial (final) von Neumann entropies of the system. 
 
For the protocol in Fig.~\ref{schematic} we have $S^V_i=0$ since the initial state is pure and $S^V_f=S^E_f$ since the final state is thermal. 
Here  $S^E_i (S^E_f)$ are  the {\it energy} entropies for the initial (final) state.  
The total entropic change $S^V_f-S^V_i$ can now be decomposed into a classical contribution $S^E_f-S^E_i$ and a quantum contribution $S^E_i-S^V_i=D(\rho||\rho_\mathrm{ed})$ arising from coherence, Eq.~\eqref{SeSv}.  The work then consists of two parts \cite{kammerlander2016, mohammady2020,francica2020}, a quantum part as defined in Eq.~\eqref{Wrelen}, as well as a classical part:
\begin{equation}\label{W}
\begin{split}
W_\mathrm{quant}=&\beta^{-1}(S^E_i-S^V_i) \\ 
W_\mathrm{class}=&U_i-U_f-\beta^{-1}(S^E_i-S^E_f).
\end{split}
\end{equation}
Note that a projection of the initial state Eq.~\eqref{rhoC} onto the energy eigenbasis prior to work extraction, i.e. $\rho \to \rho_\mathrm{ed}$, would result in $W_\mathrm{quant}=0$, but would not change $W_\mathrm{class}$. The protocol in Fig.~\ref{schematic} avoids such projection, allowing $W_\mathrm{quant}$ to be extracted.

\subsection{Classical and quantum contributions to the work extracted from a GCS}
With an initial state~\eqref{rhoC}, the outputs $W_\mathrm{quant}$ and $W_\mathrm{class}$ in Eq.~\eqref{W} can be calculated analytically. We choose $|\alpha|$ so that the GCS has mean boson number equal to the thermal state,
\begin{equation}
|\alpha|^2=\langle\hat{n}\rangle_\mathrm{therm}\equiv \langle\hat{n}\rangle.
\end{equation}
Here $\hat{n}=\hat{a}^\dagger\hat{a}$ and a ``therm'' subscript denotes expectation values for state $\rho_\mathrm{therm}$,~Eq.~\eqref{rhoT}. We have checked that this choice of $|\alpha|$ minimises $W_\mathrm{class}$ to a very good approximation over the parameters explored. The energetic and entropic terms in Eq.~\eqref{W} are
\begin{equation}\label{US}
\begin{split}
U_i-U_f=&g\left(\langle\hat{n}\rangle-\langle\delta \hat{n}^2\rangle_\mathrm{therm}\right),\\
%U_f=&\frac{1}{2\beta}-\frac{\Delta+g}{2}\langle\hat{n}\rangle\\
S^E_i=&\frac{1}{2}+\frac{1}{2}\ln \left(2\pi \langle\hat{n}\rangle\right)+O\left(\frac{1}{\langle\hat{n}\rangle}\right),\\
S^E_f=&\frac{1}{2}+\frac{1}{2}\ln \left(2\pi\langle\delta \hat{n}^2\rangle_\mathrm{therm}\right)+S_\mathrm{NG},
\end{split}
\end{equation}
where $\langle \delta\hat{n}^2\rangle=\langle\hat{n}^2\rangle-\langle\hat{n}\rangle^2$. We have also introduced
\begin{equation}
S_\mathrm{NG}=\ln\frac{1+\operatorname{erf} u}{2}-uG(u)-\ln\left[1-2uG(u)-2G(u)^2\right],
\end{equation}
with $u=\frac{\beta(g-\Delta_0)}{2\sqrt{\beta g}}$ and $G(u)=\frac{e^{-u^2}}{\sqrt{\pi}(1+\operatorname{erf}u)}=O(e^{-\beta\Delta_0^2/(4g)})$. The entropic correction $S_\mathrm{NG}$ arises since the thermal distribution is not a perfect Gaussian, and is only important for $u\lesssim 1$.

\begin{figure*}
\includegraphics[trim=0cm 5.5cm 0cm 5.5cm,clip=true,width=0.75\textwidth]{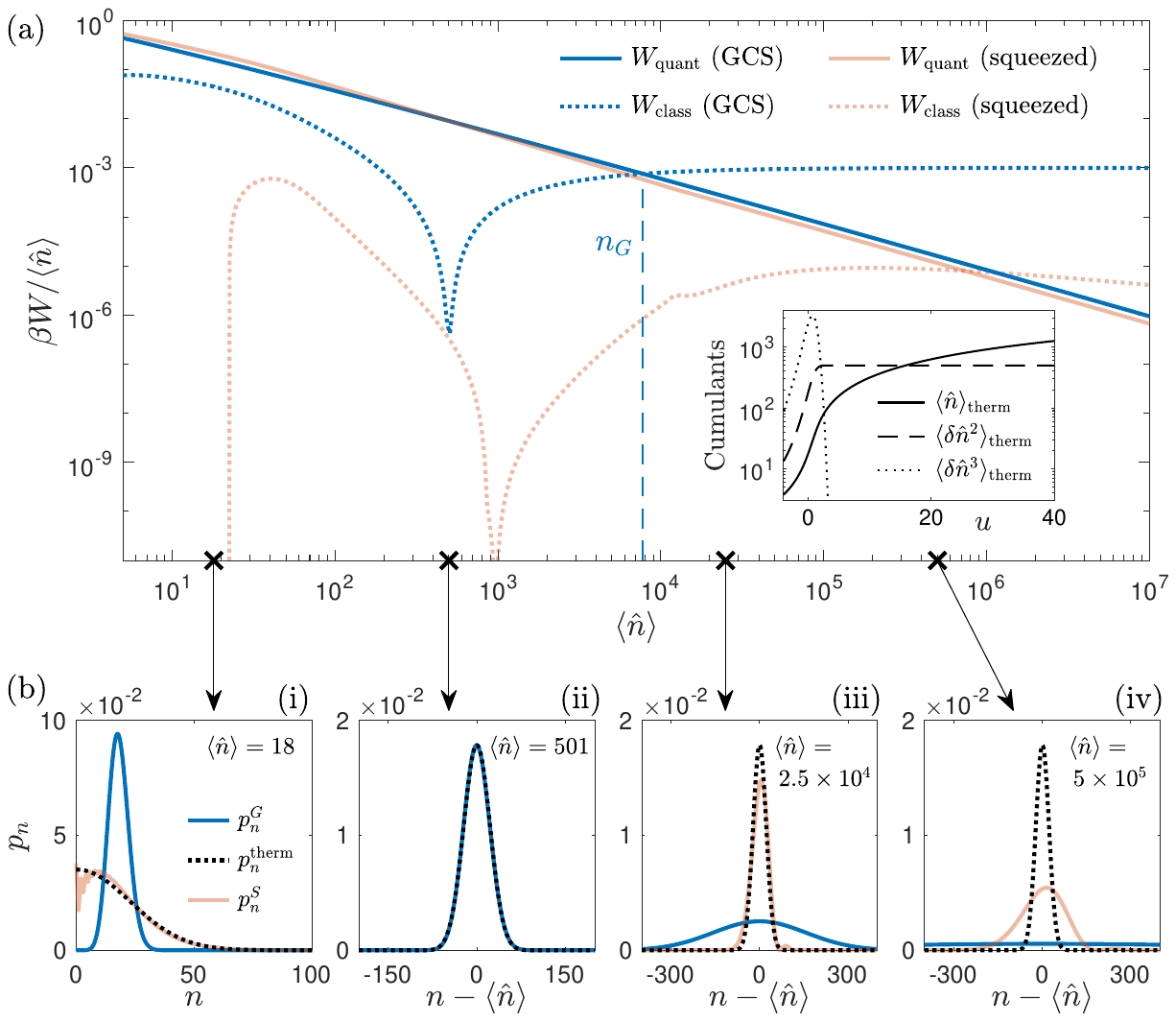}
\caption{\label{W1} (a) Work extracted from a two-mode BEC undergoing the protocol in Fig.~\ref{schematic} for different mean bosonic numbers $\langle\hat{n}\rangle$. Work extracted for an initial GCS (blue curves) divided into quantum $\beta W_\mathrm{quant}$ (blue solid curve) and classical $\beta W_\mathrm{class}$ (blue dotted curve) contributions, see Eq.~\eqref{W}. The work output is dominated by $W_\mathrm{quant}$ for $\langle\hat{n}\rangle\lesssim n_G$, with $n_G$ given by Eq.~\eqref{nc} and marked by a vertical dashed line. The number state distributions for the GCS and thermal states coincide almost perfectly when $\langle\hat{n}\rangle=1/(2\beta g)$, at which point $W_\mathrm{class}\approx 0$ is minimised (dip in blue dotted curve). Inset: First three cumulants of the thermal boson number distribution, with $\langle\delta\hat{n}^k\rangle=\langle(\hat{n}-\langle\hat{n}\rangle)^k\rangle$ for $k=2,3$. The distribution is Gaussian for $u\gg 1$, in which case $\langle\hat{n}\rangle_\mathrm{therm}\approx u/\sqrt{\beta g}$, $\langle\delta\hat{n}^2\rangle_\mathrm{therm}\approx 1/(2\beta g)$ and higher order cumulants are zero. (b) (i)-(iv) Number state distributions for different mean bosonic numbers (four panel values are marked by crosses in (a)) for a GCS (blue curve) and a thermal state (black dashed curve). The GCS distribution is narrower (wider) than the thermal distribution for $\langle\hat{n}\rangle$ smaller (larger) than $\langle\delta\hat{n}^2\rangle_\mathrm{therm}\approx 1/(2\beta g)\approx 501$, which results in non-zero $W_\mathrm{class}$. Pale red curves in (a) and (b): as for the GCS but for a squeezed state. These are included here for comparison and will be discussed in Sec.~\ref{squeeze}. The GCS and squeezed distributions coincide in (ii).  All results are for $\beta g=10^{-3}$. Different $\langle\hat{n}\rangle$ correspond to different $\beta\Delta_0$, which in (b) are: (i) $-\beta\Delta_0=0$, (ii) $-\beta\Delta_0=1$, (iii) $-\beta\Delta_0=50$, (iv) $-\beta\Delta_0=10^3$.}
\end{figure*}

The thermal mean and variance are
\begin{equation}\label{ntherm}
\langle\hat{n}\rangle_\mathrm{therm}=\frac{1}{\sqrt{\beta g}}\left(u+G(u)\right),
\end{equation}
\begin{equation}\label{ndelttherm}
\langle\delta\hat{n}^2\rangle_\mathrm{therm}=\frac{1}{2\beta g}\left(1-2uG(u)-2G(u)^2\right).
\end{equation}
Again, the $G(u)$ terms are non-Gaussian corrections, which are only important for $u\lesssim 1$. Note for $u\gg 1$ we have $u^2\approx \langle \hat{n}\rangle_\mathrm{therm}^2/(2\langle\delta\hat{n}^2\rangle_\mathrm{therm})$ and hence $u^{-1}$ is essentially the normalized second-order correlation function of the thermal bosonic field~\cite{glauber1963}. The thermal mean, variance and third central moment $\langle\delta\hat{n}^3\rangle=\langle(\hat{n}-\langle\hat{n}\rangle)^3\rangle$ are shown in the inset to Fig.~\ref{W1}(a). We focus on cases $\langle\hat{n}\rangle_\mathrm{therm}\gtrsim 1$, which is ensured when the initial detuning is less than the reservoir temperature,
\begin{equation}\label{deltamin}
-\infty<\beta \Delta_0\lesssim 1.
\end{equation}
We hence allow $\beta\Delta_0$ to be negative. This is easy to engineer in a two-mode BEC but difficult to engineer in other bosonic systems that do not have a constrained particle number, e.g.\ photons.

Analytic expressions for $W_\mathrm{class}$ and $W_\mathrm{quant}$ follow from Eq.~\eqref{W}--\eqref{ndelttherm} and are plotted in Fig.~\ref{W1}(a) as a function of $\langle\hat{n}\rangle$. Note $\langle\hat{n}\rangle$ is itself a function of $\Delta_0$, according to Eq.~\eqref{ntherm}, and hence varying $\langle\hat{n}\rangle$ corresponds to an implicit variation of $\Delta_0$. The minimum of the classical work occurs when $p_n^G$ and $p_n^\mathrm{therm}$ are as close as possible, see Fig.~\ref{W1}(b). For large $\langle\hat{n}\rangle$ both distributions are approximately Gaussian and hence will be equal when their respective means and variances are equal. Since $p_n^G$ has equal mean and variance, this requires
\begin{equation}\label{GCSconstraint}
|\alpha|^2=\langle\hat{n}\rangle_\mathrm{therm}=\langle\delta\hat{n}^2\rangle_\mathrm{therm}.
\end{equation}
This occurs when $\beta\Delta_0\approx -1$, at which point $\langle\hat{n}\rangle_\mathrm{therm}=\langle\delta\hat{n}^2\rangle_\mathrm{therm}\approx 1/(2\beta g)$. At this optimal point, $W_\mathrm{class}=O(\langle\hat{n}\rangle^{-1})$ and the work output almost exclusively consists of $W_\mathrm{quant}$. (Note our choice of units: plotting work in units of $\beta^{-1}$ is equivalent to plotting the information lost if the work was dissipated as heat, according to Landauer's principle~\cite{landauer1961}.)

For $\langle\hat{n}\rangle\ne \langle\delta\hat{n}^2\rangle_\mathrm{therm}$, the protocol is suboptimal in the sense that $W_\mathrm{class}$ is no longer approximately zero. Here the thermal bosons are either bunched or anti-bunched, caused by the boson interactions~\cite{yang2011}. For $\langle\hat{n}\rangle>\langle\delta\hat{n}^2\rangle_\mathrm{therm}$ (thermal anti-bunching), large number fluctuations in the initial state result in a large $U_i-U_f$, see Fig.~\ref{W1}(b). In this regime, $u\gg 1$ and Eq.~\eqref{US}--\eqref{ndelttherm} simplify to give
\begin{equation}\label{Wgauss}
\begin{split}
W_\mathrm{class}&\approx g\langle\hat{n}\rangle-\frac{1}{2\beta}-\frac{1}{2\beta}\ln\left(2\beta g\langle\hat{n}\rangle \right),\\
W_\mathrm{quant}&\approx \frac{1}{2\beta}+\frac{1}{2\beta}\ln\left(2\pi\langle\hat{n}\rangle\right).
\end{split}
\hspace{1cm}(u\gg 1)
\end{equation}
Using Eq.~\eqref{Wgauss}, we find that $W_\mathrm{quant}>W_\mathrm{class}$ for
\begin{equation}\label{nc}
\langle\hat{n}\rangle\lesssim \frac{1}{\beta g}\left[-\Omega_{-1}\left(-\sqrt{\frac{\beta g}{4\pi \exp(2)}}\right)\right]\equiv n_G,
\end{equation}
see Fig.~\ref{W1}(a). Here $\Omega_k$ is the $k$th branch of the omega function~\footnote{Also known as the Lambert $W$-function, denoted by $W_k$. We avoid this notation to avoid confusion with work output.}. The function $-\Omega_{-1}(-x)$ satisfies $1\le -\Omega_{-1}(-x)<\infty$ for $x\in (0,\exp(-1)]$ and monotonically decreases with increasing $x$, with $-\Omega_{-1}(-x)\sim \ln \left(\frac{1}{x}\ln \frac{1}{x}\right)$ for small $x$~\cite{corless1996}. Hence, ignoring logarithmic corrections, $n_G$ decreases with increasing $\beta g$ as $n_G\sim 1/(\beta g)$. For $\langle\hat{n}\rangle>n_G$, the classical work output grows extensively due to the extensive growth of $U_i-U_f$, see Eq.~\eqref{Wgauss}. The quantum work also increases with $\langle\hat{n}\rangle$ due to increased coherence in the initial state. The increase, however, is subextensive, with $W_\mathrm{quant}\sim \frac{1}{2\beta}\ln \langle\hat{n}\rangle$. For $\langle\hat{n}\rangle<\langle\delta\hat{n}^2\rangle_\mathrm{therm}$ (thermal bunching), the number fluctuations of the initial state are less than the final state, see Fig.~\ref{W1}(b). Although this increases $W_\mathrm{class}$, we still have $W_\mathrm{quant}>W_\mathrm{class}$ as long as $\langle\hat{n}\rangle\gg 1$, see Fig.~\ref{W1}(a).

\subsection{Improved protocol using squeezing}\label{squeeze}
The boson number distribution of a GCS is Poissonian, and hence the mean and variance are always equal. For sufficiently large mean boson number, Eq.~\eqref{nc}, the work output becomes dominated by $W_\mathrm{class}$. The classical work output can be suppressed much more effectively by using a squeezed initial state. Squeezing of a two-mode BEC is possible using multiple techniques~\cite{ma2011,pezze2018}, including time evolution with a one-axis twisting Hamiltonian~\cite{kitagawa1993,wang2003,esteve2008} and parametric downconversion via spin-changing collisions~\cite{duan2000,pu2000,gross2011}. A displaced squeezed state is~\cite{stoler1970,yuen1976,caves1981},
\begin{equation}\label{rhoS}
\ket{\alpha,\zeta}=\hat{D}(\alpha)\hat{S}(\zeta)\ket{0},
\end{equation}
where $\hat{S}(\zeta)=e^{(\zeta^* \hat{a}\hat{a}-\zeta \hat{a}^\dagger\hat{a}^\dagger)/2}$ is the squeezing operator and $\zeta=|\zeta|e^{i\theta}$ parameterises the squeezing. The distribution $p_n^S=|\braket{n|\alpha,\zeta}|^2$ is~\cite{gerry2005,gong1990}
\begin{equation}\label{pnSqueezed}
p_n^S=\frac{(\frac{1}{2}\tanh |\zeta|)^n}{n!\cosh |\zeta|}\left|H_n(\lambda\alpha)\right|^2e^{-\alpha^2(1+\cos\theta\tanh |\zeta|)},
\end{equation}
with $H_n$ the Hermite polynomials ($H_0(x)=1$, $H_1(x)=2x$, etc.) and
\begin{equation}
\lambda=\frac{e^{-i\theta/2}\cosh |\zeta|+e^{i\theta/2}\sinh |\zeta|}{\sqrt{\sinh 2|\zeta|}}.
\end{equation}

\begin{figure*}
\includegraphics[trim=0cm 4.5cm 0cm 5.5cm,clip=true,width=0.75\textwidth]{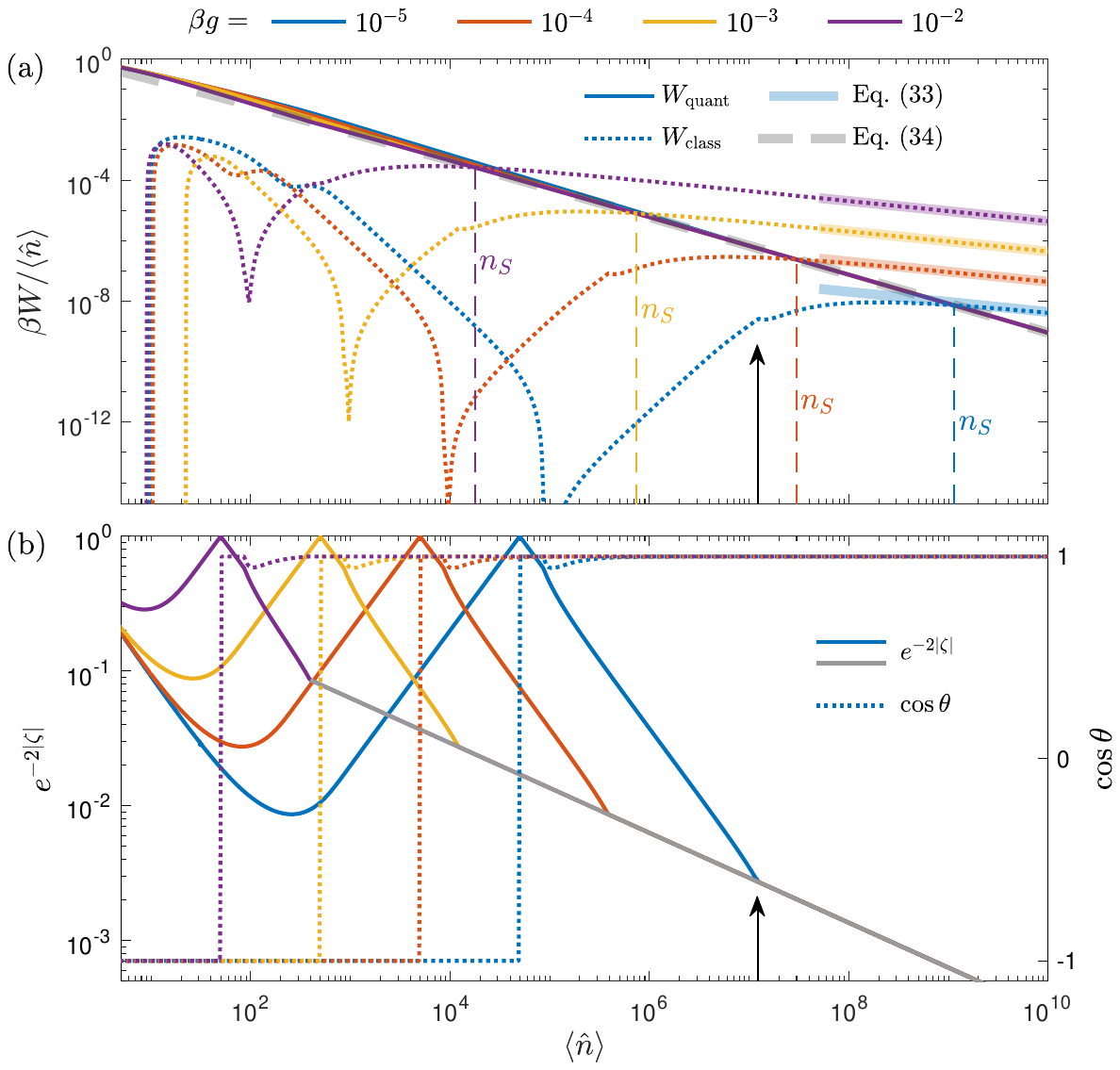}
\caption{\label{Wsqueeze} (a) Work extracted from an initial squeezed state, divided into quantum $\beta W_\mathrm{quant}$ (solid curves) and classical $\beta W_\mathrm{class}$ (dotted curves) contributions. Colours indicate different values of $\beta g$. The work output is dominated by $W_\mathrm{quant}$ for $\langle\hat{n}\rangle\lesssim n_S$, with $n_S$ given by Eq.~\eqref{ns} and marked by vertical dashed lines. Lighter-coloured thicker lines give the large $\langle\hat{n}\rangle$ result for $W_\mathrm{class}$, Eq.~\eqref{Wclargen}, demonstrating subextensive scaling of $W_\mathrm{class}$. Grey dashed line gives the large $\langle\hat{n}\rangle$ result for $W_\mathrm{quant}$, Eq.~\eqref{Wqlargen}. (b) Optimised squeezing parameters $e^{-2|\zeta|}$ (solid curves, left axis) and $\cos\theta$ (dotted curves, right axis) for the squeezed state (see text for details). For $\langle\hat{n}\rangle \le 2(3\beta g)^{-3/2}$ the squeezed variance can be matched to the thermal variance with appropriate choice of $e^{-2|\zeta|}$. For larger $\langle\hat{n}\rangle$ matching is no longer possible, see Eq.~\eqref{squeezeconstraint}; minimising the squeezed variance then gives $e^{-2|\zeta|}=(4\langle\hat{n}\rangle)^{-1/3}$ (gray line), independent of $\beta g$. This results in a discontinuity in the slope of $e^{-2|\zeta|}$ and $W_\mathrm{class}$ at $\langle\hat{n}\rangle=2(3\beta g)^{-3/2}$ (marked by vertical arrows for $\beta g=10^{-5}$).}
\end{figure*}

The boson number distribution for a squeezed state can have a variance smaller or larger than its mean, controllable via the squeezing parameter $\zeta$. This additional control allows the initial number distribution to be matched more closely to $p_n^\mathrm{therm}$, see Fig.~\ref{W1}(b). The work output from an initial squeezed state is compared with that of a GCS in Fig.~\ref{W1}(a). Squeezing suppresses $W_\mathrm{class}$ over a much larger range of $\langle\hat{n}\rangle$ compared to the GCS. Note the small but finite minimum of $W_\mathrm{class}/\langle\hat{n}\rangle$ for a GCS, which arises due to the approximation in Eq.~\eqref{png}, is reduced even further by squeezing, see Fig.~\ref{W1}(a). Further squeezed results are shown in Fig.~\ref{Wsqueeze}(a) for different values of $\beta g$. For the squeezed state, we choose $|\alpha|^2$ so that the mean boson number is equal to that of the thermal state and choose $\zeta$ to minimise differences in higher order moments. Details are discussed in Appendix~\ref{squeezeParams} and the values of $|\zeta|$ and $\cos\theta$ obtained are plotted in Fig.~\ref{Wsqueeze}(b). Numerical details to obtain $p_n^S$ are described in Appendix~\ref{squeezeNum}.

Although squeezing gives much greater control over the initial boson number distribution, its variance is still constrained by $\langle\hat{n}\rangle$~\cite{bondurant1984}. For large $\langle\hat{n}\rangle$ the minimum variance possible for a squeezed state is (see Appendix~\ref{squeezeVar})
\begin{equation}\label{variance}
\operatorname{min}\langle \delta \hat{n}^2\rangle\approx \left(\frac{27}{32}\right)^{1/3}\langle\hat{n}\rangle^{2/3}
\end{equation}
Achieving $\langle\delta\hat{n}^2\rangle=\langle\delta\hat{n}^2\rangle_\mathrm{therm}$ is hence only possible when the mean boson number is not too large,
\begin{equation}\label{squeezeconstraint}
\langle\hat{n}\rangle \le \sqrt{\frac{32\langle\delta\hat{n}^2\rangle_\mathrm{therm}^3}{27}}\approx \sqrt{\frac{4}{27 (\beta g)^3}}.
\end{equation}
The constraint~\eqref{squeezeconstraint} should be contrasted with the strict constraint for a GCS, Eq.~\eqref{GCSconstraint}. Matching the mean and variances of $p_n^S$ and $p_n^\mathrm{therm}$ gives $U_i=U_f$ and hence $W_\mathrm{class}=\beta^{-1}(S_f^E-S_i^E)$, see Eq.~\eqref{W}.

If the mean boson number is too large such that Eq.~\eqref{squeezeconstraint} is not satisfied, the variance of the squeezed distribution cannot be made small enough to match $\langle\delta\hat{n}^2\rangle_\mathrm{therm}$. We then choose $\langle \delta \hat{n}^2\rangle$ to take on its minimum value Eq.~\eqref{variance} by choosing $e^{-2|\zeta|}=(4\langle\hat{n}\rangle)^{-1/3}$ (see Appendix~\ref{squeezeVar}). This results in subextensive scaling of $W_\mathrm{class}$ for large $\langle\hat{n}\rangle$,
\begin{equation}\label{Wclargen}
W_\mathrm{class}\sim g\left(\frac{27}{32}\right)^{1/3}\langle\hat{n}\rangle^{2/3},
\end{equation}
see Fig.~\ref{Wsqueeze}(a). Notably, this contrasts the extensive scaling of $W_\mathrm{class}$ for a GCS, which arises from the strict constraint $\langle\hat{n}\rangle=\langle\delta\hat{n}^2\rangle$, see Fig.~\ref{W1}(a) and Eq.~\eqref{Wgauss}. For large $\langle\hat{n}\rangle$, we find $W_\mathrm{quant}$ is well approximated by
\begin{equation}\label{Wqlargen}
W_\mathrm{quant}\approx \frac{1}{2\beta}+\frac{1}{2\beta}\ln\left[2\pi\left(\frac{27}{32}\right)^{1/3}\langle\hat{n}\rangle^{2/3}\right],
\end{equation}
see Fig.~\ref{Wsqueeze}(a). This follows from approximating $p_n^S$ by a Gaussian distribution and using Eq.~\eqref{variance}. Equations~\eqref{Wclargen} and~\eqref{Wqlargen} give the following criteria for $W_\mathrm{quant}>W_\mathrm{class}$,
\begin{equation}\label{ns}
\langle\hat{n}\rangle\lesssim \sqrt{\frac{32}{27}n_G^3}\equiv n_S,
\end{equation}
with $n_G$ the equivalent bound for a GCS given by Eq.~\eqref{nc}. The estimate Eq.~\eqref{ns} agrees well with our numerics, see Fig.~\ref{Wsqueeze}(a). Notably, $n_S\gg n_G$ for large $n_G$.

For small $\langle\hat{n}\rangle$, we observe an abrupt drop in $W_\mathrm{class}$, see Fig.~\ref{Wsqueeze}(a). At this point, $S_i^E$ increases above $S_f^E$ and hence $W_\mathrm{class}$ becomes negative ($W_\mathrm{class}=S_f^E-S_i^E$ for small $\langle\hat{n}\rangle$, see Eq.~\eqref{squeezeconstraint}). We expect that the increase in $S_i^E$ above $S_f^E$ is due to oscillations that arise in the squeezed distribution for small $\langle\hat{n}\rangle$~\cite{schleich1987,schleich1987b}, see Fig.~\ref{W1}(b)~\footnote{The change of sign of $W_\mathrm{class}$ means that there is value of $\langle\hat{n}\rangle$ where $W_\mathrm{class}=0$ despite differences in the squeezed and thermal distributions. The choice of squeezing parameters can possibly be optimised to find similar ``serendipitous'' points of zero or small $W_\mathrm{class}$ for other values of $\langle\hat{n}\rangle$.}.

\subsection{Relation with entanglement}
Finally, we show how the work from coherence in this system can equivalently be interpreted as work from entanglement~\cite{perarnau2015,sapienza2019,touil2021,francica2017}. To see this, we evaluate the entanglement entropy $\mathcal{S}$ between the two modes of the BEC. A sufficient condition for entanglement is then $\mathcal{S}>0$~\cite{bennett1996}. The entanglement entropy of a pure state $\rho$ is
\begin{equation}\label{entanglementS}
\mathcal{S}=-\operatorname{Tr}_1\rho_1\log \rho_1,
\end{equation}
where $\rho_1=\operatorname{Tr}_2\rho$ and $\operatorname{Tr}_i$ is a partial trace over boson states of modes $i=1,2$.   The trace $\operatorname{Tr}_2$ can be evaluated in the number state basis for both modes. Due to conservation of total particle number, this projects $\rho$ onto the number state basis for the first mode,
\begin{equation}\label{rho1}
\rho_1=\sum_{n=0}^N\braket{n,N-n|\rho|n,N-n}\ket{n}_1\prescript{}{1}{\bra{n}}
\end{equation}
with $\ket{n}_1\prescript{}{1}{\bra{n}}=\operatorname{Tr}_2 \ket{n,N-n}\bra{n,N-n}$. The ``$1$'' subscript distinguishes $\ket{n}_1$ from the implicitly two-mode state $\ket{n}$. Equations~\eqref{entanglementS} and~\eqref{rho1} give
\begin{equation}
\mathcal{S}=S_i^E=\beta W_\mathrm{quant}.
\end{equation}
Hence the coherence in this system arises due to entangling correlations between the two modes, which is subsequently extracted as $W_\mathrm{quant}$. The final thermal state is a classical mixture of product states, and hence does not possess any entanglement. The quantum work output from coherence can therefore equivalently be interpreted as work from entanglement. Squeezing results in additional entanglement according to the criteria in~\cite{sorensen2001,sorensen2001b}.

\section{Discussion and conclusion} We have shown how work can be extracted from coherence in a two-mode BEC. Work almost entirely from coherence can be extracted from a GCS with precisely tuned mean boson number. The protocol can be substantially improved using squeezing, enabling work extraction predominantly from coherence over a much broader range of parameters. Due to the many-body nature of the system, the work from coherence can equivalently be interpreted as work from entanglement.

The coherent steps in our protocol can be realised in current experiments with two-mode BECs, which could be spatial or spin modes~\cite{leggett2001}. For two spatial modes the detuning can be controlled by varying trap depth~\cite{albiez2005} and populations can be coherently controlled using Josephson oscillations~\cite{albiez2005,gati2007}. For two spin states the detuning can be controlled using Zeeman fields~\cite{matthews1998,hall1998} and populations can be controlled using a Rabi pulse~\cite{matthews1999,zibold2010}. The effect of changing $\Delta$ in the protocol could also be achieved by manipulating $g$ using a Feshbach resonance~\cite{papp2008,tojo2010,chin2010}. Squeezing of a two-mode BEC can be realised using a variety of techniques~\cite{ma2011,pezze2018,kitagawa1993,wang2003,esteve2008,duan2000,pu2000,gross2011}. Our protocol requires $\beta g\ll 1$, which is satisfied in most realisations of dilute gaseous BECs~\cite{ketterle1999,dalfovo1999}.

The experimentally challenging part of the protocol is the thermalisation of the system with a reservoir (step 3 in Fig. 1). This could potentially be achieved by immersing the system in a thermal quantum gas of another species, in analogy with sympathetic cooling~\cite{modugno2001}. An optical tweezer trap would allow the system to be moved in and out of the reservoir as needed. Particle exchange between the two modes during thermalization could occur via tunnelling for two spatial modes~\cite{grabert1985,fisher1985} or via spin-changing collisions with the reservoir for two spin states~\cite{bradley2014,Kawaguchi2012R,StamperKurn2013a,bouton2021}. We have assumed decoherence only occurs during the thermalising step of our protocol. Additional decoherence will likely reduce the work output by a term that scales linearly with the irreversible entropy production~\cite{mohammady2020}.

The Hamiltonian Eq.~\eqref{H} also describes a large, fully-connected spin chain~\cite{ulyanov1992} and hence could be realised in other atomic systems~\cite{pezze2018}, including atomic ensembles coupled to light~\cite{hammerer2010,leroux2010}, Rydberg arrays~\cite{bouchoule2002,saffman2010} and trapped ions~\cite{sorensen2000,meyer2001}. The Hamiltonian also resembles that of photons in a non-linear medium~\cite{gerry2005}, with the important difference that we allow the free-photon spectrum to be negative, $\Delta<0$. To realise this, photon number would need to be conserved~\cite{klaers2010a,klaers2010b}.

We have limited our analysis to Glauber coherent and squeezed states. Improved operation may be possible using self-phase modulation of the squeezed state~\cite{kitagawa1986} or by using non-Gaussian states~\cite{walschaers2021} at the expense of more involved state generation. We have also limited our analysis to initial states with zero von Neumann entropy; relaxing this provides an interesting area for future investigation. Extracting work from coherence can enhance power output compared to classical systems~\cite{uzdin2015}. Our results provide a starting point to explore this in a two-mode BEC, and may result in a many-body quantum advantage when incorporated into an engine cycle.

\section*{Data Availability}
The code used to produce the data shown in Figs.~\ref{W1} and~\ref{Wsqueeze} is available upon reasonable request to LW, lewis.williamson@uq.edu.au.

\section*{Acknowledgements}
This research was supported by the Australian Research Council Centre of Excellence for Engineered Quantum Systems (EQUS, CE170100009) and the Australian government Department of Industry, Science, and Resources via the Australia-India Strategic Research Fund (AIRXIV000025). FC and JA gratefully acknowledge funding from the Foundational Questions Institute Fund (FQXi-IAF19-01) and EPSRC (EP/R045577/1). JA thanks the Royal Society for support.

\appendix
\section{Choosing squeezing parameters}\label{squeezeParams}
We choose $\alpha$ and $\zeta$ for the initial squeezed state as follows. The first three cumulants of $p_n^S$ are evaluated from the generating function $K(s)=\ln\sum_n p_n^S e^{sn}$, which can be written in closed form using Mehler's formula~\cite{erdelyi1953}
\begin{equation}
\sum_{n=0}^\infty\frac{H_n(x)H_n(y)}{n!}\left(\frac{u}{2}\right)^n=\frac{1}{\sqrt{1-u^2}}e^{\frac{2xyu-(x^2+y^2)u^2}{1-u^2}}.
\end{equation}
This gives 
\begin{equation}\label{cumulants}
\begin{split}
\langle \hat{n}\rangle=&\sinh^2 |\zeta|+|\alpha|^2,\\
\langle \delta \hat{n}^2\rangle=&2\cosh^2 |\zeta|\sinh^2 |\zeta|+|\alpha|^2\left[\cosh \left(2|\zeta|\right)-\sinh \left(2|\zeta|\right)\cos\theta\right],\\
\langle \delta n^3\rangle=&\cosh \left(2|\zeta|\right)\sinh^2 \left(2|\zeta|\right)\\
&+\frac{3|\alpha|^2}{2}\left[\cosh \left(4|\zeta|\right)-\frac{1}{3}-\sinh \left(4|\zeta|\right)\cos\theta\right],
\end{split}
\end{equation}
with $\langle \delta n^k\rangle=\langle (n-\langle \hat{n}\rangle)^k\rangle$ for $k=2,3$. We set $\langle \hat{n}\rangle=\langle\hat{n}\rangle_\mathrm{therm}$ by choosing $|\alpha|^2=\langle\hat{n}\rangle_\mathrm{therm}-\sinh^2|\zeta|$. We solve for $|\zeta|$ by minimising $|\langle\delta n^2\rangle-\langle \delta \hat{n}^2\rangle_\mathrm{therm}|$. Since the Hamiltonian~\eqref{H} depends only on $\langle \hat{n}\rangle$ and $\langle \delta\hat{n}^2\rangle$, this choice minimises $|U_f-U_i|$. Additionally, for not-too-large squeezing this minimises $|S^E_f-S^E_i|$. We then choose $\cos\theta$ to minimise $|\langle \delta n^3\rangle|$.

\section{Numerical details for extracting squeezed state statistics}\label{squeezeNum}
Numerical computation of $p_n^S$ requires care for large $n$ due to numerical overflow and underflow~\cite{bunck2009}. We evaluate $p_n^S$ using the underlying recursion relation~\cite{gerry2005,gong1990} $c_n=\braket{n|\alpha,\zeta}=\frac{\alpha(1+e^{i\theta}\tanh|\zeta|)}{\sqrt{n}}c_{n-1}-e^{i\theta}\tanh|\zeta|\sqrt{\frac{n-1}{n}}c_{n-2}$. After each iteration we normalise the distribution. Defining $\Delta n=\operatorname{ceil}(50\sqrt{\langle \delta \hat{n}^2\rangle})$, we begin the recursion at $n^*=\operatorname{max}(0,\operatorname{round}(\langle \hat{n}\rangle-\Delta n/2))$ with $c_{n^*}=1$ and $c_{n<n^*}=0$, and evaluate terms up to $n^*+\Delta n$. In cases where $n^*>0$, we find $|\alpha|\gg 1$ and hence $c_{n^*+1}\approx \frac{\alpha(1+e^{i\theta}\tanh|\zeta|)}{\sqrt{n}}c_{n^*}$ (this is exact for $n^*=0$). Hence why the values of $c_{n<n^*}$ are not needed.

\section{Minimum variance of a squeezed state}\label{squeezeVar}
To a very good approximation $\cos\theta=\pm 1$, see Fig.~\ref{Wsqueeze}(b). Defining $x=e^{2|\zeta|\cos\theta}\left(=e^{\pm 2|\zeta|}\right)$, the equation for $\langle \delta \hat{n}^2\rangle$ is (see Eq.~\eqref{cumulants}),
\begin{equation}\label{deltan}
\langle\delta\hat{n}^2\rangle=\frac{1}{8 x^2}\left(x^4-4 x^2+8\left(\langle\hat{n}\rangle+\frac{1}{2}\right)x-1\right).
\end{equation}
Stationary points of $d\langle\delta\hat{n}^2\rangle/dx$ satisfy the polynomial
\begin{equation}\label{polynomial}
x^4-4\left(\langle\hat{n}\rangle+\frac{1}{2}\right)x+1=0.
\end{equation}
Equation~\eqref{polynomial} can be solved for $x$ using standard methods. For large $\langle\hat{n}\rangle$ this gives two real roots and two complex roots. The two real roots are
\begin{equation}
\begin{split}
x_1\approx &\frac{1}{4\langle\hat{n}\rangle+2},\\
x_2\approx &(4\langle\hat{n}\rangle)^{1/3},
\end{split}
\end{equation}
with $x_1$ corresponding to a local maximum of Eq.~\eqref{deltan} and $x_2>x_1$ a local minimum. Substituting $x\le x_1$ into the expression for $\langle\hat{n}\rangle$ (Eq.~\eqref{cumulants}) would give a negative value for $|\alpha|^2$, hence $x> x_1$. Therefore $x_2$ gives the global minimum of $\langle\delta\hat{n}^2\rangle$ for valid values of $x$. Substituting $x_2$ into Eq.~\eqref{deltan} gives Eq.~\eqref{variance} for large $\langle\hat{n}\rangle$.\\
\\
\textit{For the purpose of open access, the authors have applied a `Creative Commons Attribution' (CC BY) licence to any Author Accepted Manuscript version arising from this submission.}

\end{document}